\documentclass[12pt]{iopart}

\usepackage{citesort}
\usepackage{gensymb}
\usepackage{graphicx}
\usepackage{epstopdf}

\usepackage{iopams}
\usepackage[utf8]{inputenc}

\begin{document}

\title{LCLS in - photon out: fluorescence measurement of neon using soft x-rays}

\author{Razib Obaid $^1$, Christian Buth $^2$, Georgi L. Dakovski $^5$, Randolf Beerwerth $^{3, 4}$, Michael Holmes $^5$, Jeff Aldrich $^5$, Ming-Fu Lin $^5$, Michael Minitti $^5$, Timur Osipov $^5$, William Schlotter $^5$, Lorenz S. Cederbaum $^2$, Stephan Fritzsche $^{3, 4}$, and Nora Berrah $^1$}

\vspace{1pc}
\address{$^1$ Department of Physics, University of Connecticut, Storrs, Connecticut, USA.}

\address{$^2$ Theoretische Chemie, Physikalisch-Chemisches Institut, Ruprecht-Karls-Universität Heidelberg, Heidelberg, Germany.}

\address{$^3$ Helmholtz-Institut Jena, Jena, Germany.}

\address{$^4$ Theoretisch-Physikalisches Institut, Friedrich-Schiller-Universität Jena, Jena, Germany.}

\address{$^5$ SLAC National Accelerator Laboratory, 2575 Sand Hill Road, Menlo Park, California, 94025, USA.}

\ead{razib.obaid@uconn.edu}
\vspace{10pt}
\begin{indented}
\item[]July 2017
\end{indented}

\begin{abstract}

We measured the fluorescence photon yield of neon upon soft x-ray ionization ($\sim$1200 eV) from the x-ray free electron laser at Linac Coherent Light Source, and demonstrated the usage of a grazing incidence spectrometer with a variable line-spacing grating to perform x-ray fluorescence spectroscopy on a gas phase system. Our measurements also allowed us to estimate the focal size of the beam from the theoretical description developed, in terms of the rate equation approximation accounting for photoionization shake off of neutral neon and double Auger decay of single core holes.

\end{abstract}


\section{Introduction}
\begin{figure*}[ht]
	\centering
	\includegraphics[width = 0.9 \textwidth, scale = 0.9 ]{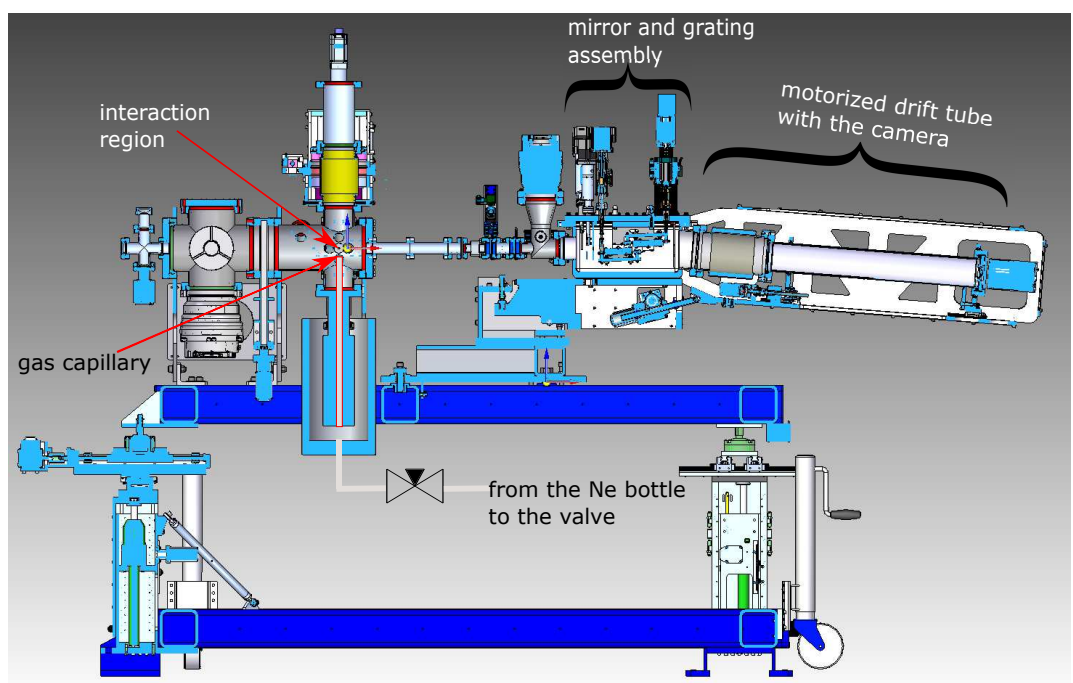}
	\caption{Vertical slice of the Liquid Jet Endstation showing the chamber with the variable line spacing grating spectrometer and the gas capillary.}
	\label{vls_setup}
\end{figure*} 

Free-electron lasers (FEL) such as the Linac Coherent Light Source (LCLS) at SLAC National Accelerator Laboratory \cite{emma,arthurJ}, SPring-8 Angstrom Compact FEL (SACLA) at Spring-8 \cite{ishikawa}, and XFEL at DESY \cite{altarelli} have ushered an era of understanding light-matter interaction using x-ray pulses with unprecedented intensity ($>$ 10$^{18}$ W/cm$^2$) and short pulse duration \cite{berrah,bostedt}. The response of such excitation in atoms and molecules reveal the underlying femtosecond dynamics that have not been observed before. Through multi-photon ionization of gas phase isolated atoms and molecules by x-rays, and studying the ionization product (ions and electrons), the structure and underlying photon induced dynamics has been studied at LCLS. One such target atom, neon, has been extensively examined as a model system for ionization and damage threshold studies \cite{youngL,doumy} where all 10 electrons were stripped off the atom through processes including inner-shell ionization leading to Auger electron removal. However, ion and electron yield spectroscopy provides only one side of the picture of the interaction with x-rays, while fluorescence provides the complementary view on the atomic system undergoing transitions following specific selection rules. Previously, there was an experiment \cite{tamasaku} at SACLA FEL investigating double core holes in krypton by measuring the fluorescence yield with hard x-rays (15 keV) ionization. Additionally, fluorescence spectra also has been shown to be useful in examining any competing mechanisms, such as x-ray induced Rabi oscillations under strong FEL field with neon \cite{cavaletto}.

In this work, we report our investigation of the neon produced fluorescence subsequent to ionization with soft x-rays of about 1200 eV utilizing a Variable Line Spacing (VLS) grating spectrometer at the Soft X-ray Materials Science Research (SXR) beamline \cite{schlotter} at LCLS. Irrespective of the low target density obtained using a capillary and the large focal area of the beam, the photon flux was sufficient enough to produce fluorescence count of ~1.2 photons/shot within the acceptance angle of the spectrometer. Three distinct peaks result for neon following photoionization shake-off and Auger decay. Using the theory \cite{buth_ne}, we provide the identification of the peaks resulting thereof. Our measurements confirm two of the peaks resulting from 1s$^{1}$2s$^{2}$2p$^{6}$ $\rightarrow$ 1s$^2$2s$^2$2p$^5$ , and a combination of 1s$^1$2s$^2$2p$^5$ $\rightarrow$ 1s$^2$2s$^2$2p$^4$ and 1s$^1$2s$^1$2p$^6$ $\rightarrow$ 1s$^2$2s$^1$2p$^5$ transitions, identified in earlier experiments measuring K emission spectrum of neon \cite{keski-rekhonen}. We also identified the third peak originating from the fluorescence after a core ionization – Auger decay – core ionization (CAC) process, which is a two photon initiated channel that cannot be accessed with a synchrotron source. Furthermore, we were able to use the theory to estimate the focal spot size of $7.30_{-0.77}^{+1.09} \times 7.30_{-0.77}^{+1.09}$  $\mu$m$^2$ of the x-ray beam which is usually not very well known with most type of FEL experiments. All theoretical computations associated are provided in the supplementary data.

\section{Experiment}
\begin{figure*}[ht]
	\centering
	\includegraphics[width = 1.0 \textwidth, scale = 1.0 ]{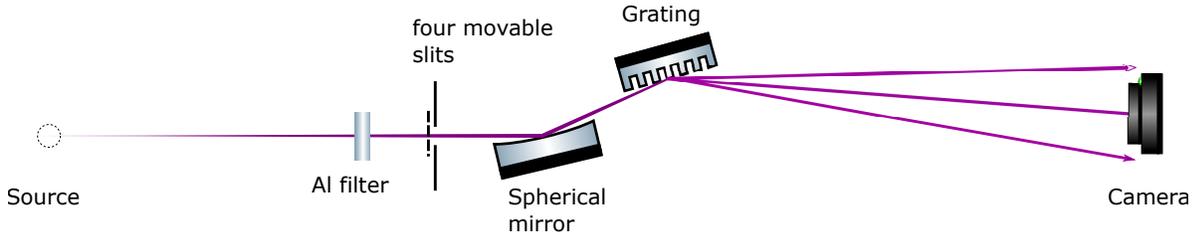}
	\caption{Conceptual layout of the VLS spectrometer with the key components.}
	\label{spectrometer_layout}
\end{figure*} 

The experiment was performed at the SXR beamline at LCLS. The beam from the FEL was focused by a Kirkpatrick-Baez mirror in the Liquid Jet Endstation chamber fitted with a VLS spectrometer as shown in Figure \ref{vls_setup}. The neon gas was introduced into the chamber through a capillary of $\sim$120 $\mu$m diameter. The base pressure of the chamber was $\sim$10$^{-9}$ Torr before introducing the gas, and, during data acquisition, the pressure elevated to 10$^{-4}$ Torr due to neon injection. The photon energy of 1203 eV, with about 6 eV bandwidth, was chosen for the ionization to be sufficiently above the K edge of neutral neon. The full width half maximum (FWHM) focal spot was estimated to be at most 20 $\times$ 20 $\mu$m$^2$, which is limited by the resolution of the imaging camera.

The working principle of a VLS spectrometer is described in detail elsewhere \cite{chuang}. The system built at LCLS is a Hettrick-Underwood spectrometer \cite{hettrick} with a combination of a spherical focusing mirror and a VLS grating of 1200 lines/mm groove density. The energy range of this grating is between 250 eV to 1 keV. Before the mirror-grating assembly, an aluminum (Al) filter of 200 nm thickness is used to remove ambient light. The four slit blades (both in vertical and horizontal axis) can be used to define the illumination profile downstream to the grating. A conceptual layout of the mirror-grating assembly is shown in Figure \ref{spectrometer_layout}. For this particular experiment, since the slits were not used, the acceptance angle for the spectrometer is estimated to be 1.2 $\times$ 10$^{-5}$ steradians, which is defined by the length and angle of the spherical mirror, and its distance from the source; and the detector width and its distance from the source. The VLS grating, with the grove density as a function of distance from the origin of the grating, ensures minimum optical aberration for a narrow energy range on a fixed detector. The advantage of this system is that for a change in photon energy from the interaction region, only the rotation of the detector is enough to track focal curve of the mirror-grating assembly. Both the mirror and the grating have a rotational resolution of 2 $\mu$rad. To scan for different order of the diffraction, the drift-tube, with the attached detector, is rotated using a NEMA 17 gearhead. The motorized drift-tube can be moved from 1$\degree$ above horizontal to 10$\degree$ below horizontal. It can also be subjected to mechanical roll ($\pm$ 0.5$\degree$ at a 10 $\mu$rad resolution) for twisting the detector position relative to the grating plane.  For the VLS spectrometer at LCLS, a commercial CCD detector (Andor, Newton 940) is used. The camera is interfaced to a 6 inch conflat flange, and is demountable so that other pixel detectors can be interfaced. The exposure time of the camera can be varied, and for this particular data run, the integration time of 600 shots/image (5 s for 120 Hz FEL repetition rate) is used.

The pixel-to-energy calibration is done using Zn and Cu solid target irradiated by 1200 eV FEL photons, which generated known peaks in the 1st diffraction order for Cu L-$\alpha$ , L-$\beta$ (929.7 eV, 949.8 eV) \cite{thompson_vaughan} and in the 2nd diffraction order for Zn L-$\alpha$, L-$\beta$ (1011.7 eV, 1034.7 eV) \cite{thompson_vaughan} emission lines. The overall spectrometer resolution was measured to be 0.75 eV. The pulse energy of LCLS was $\sim$1.2 mJ and is measured using the front end enclosure gas detector upstream of the beamline monochromator. At the interaction region, about 50\% of the pulse energy was estimated to be available for ionization corresponding to $\sim$1.6 $\times$ 10$^{12}$ photons/pulse.

\section{Theory}
We theoretically examine the ionization of neon atoms with x-rays from the LCLS and calculate the fluorescence photon yields. The details of the theoretical framework have been discussed elsewhere \cite{buth_ne}. The electronic structure of neon in all possible nonrelativistic cationic configurations is treated using a relativistic multiconfiguration approach based on Grasp2K \cite{jonsson}. We compute Auger decay, radiative decay, and photoionization cross sections with Ratip \cite{fritzsche}. Photoionization cross sections for a hydrogen-like ion are found with FAC \cite{gu}. We are interested in configuration-averaged quantities, so we performed a probabilistic average over the fine-structure-resolved results \cite{buth_ne}. To gain insights into the time-dependent absorption of x-rays and the resulting decay processes, we use the rate-equation approximation. Therefore, equations are formulated for the rate of change of the probability of an atom to be in a specific cationic configuration based on photoabsorption and decay processes. However, this does not capture the basic physics yet \cite{doumy} as there are important two-electron emission effects in neon. First, photoionization shake off amounts to 23\% above the neon K edge \cite{saito}, i.e., not only the core electron is ejected but additionally a valence electron leaves the atom. Second, double Auger decay occurs for Ne 1s holes where two electrons are emitted instead of one with an amount of 5.97\% \cite{kanngieser,hikosaka}. Consequently, we extended the rate equations to incorporate these processes \cite{buth_ne}. To predict fluorescence spectra based on rate equations, we calculate the probability for x-ray photon emission. Thereby, the emission rate is given by the probability to be in a specific configuration times the radiative decay width. Initially, the emission probability vanishes and all other quantities are known, so we can immediately integrate this equation. In principle, we need to solve the rate equations for a very long time interval for radiative transitions where no competing Auger decay is available. However, as all probability trapped in such a configuration decays radiatively, we can calculate the photon emission probability up to a chosen time when all other decays have taken place and then solve a decay equation \cite{buth_ne}. From the probability for x-ray emission, we calculate the photon yield which is the probability for photon emission on a specific transition normalized to the sum of the probabilities for photon emission on all x-ray transitions \cite{buth_ne}.

\section{Results and Discussion}
\begin{figure*}[ht]
	\includegraphics[width = 1.0 \textwidth, scale = 1.0 ]{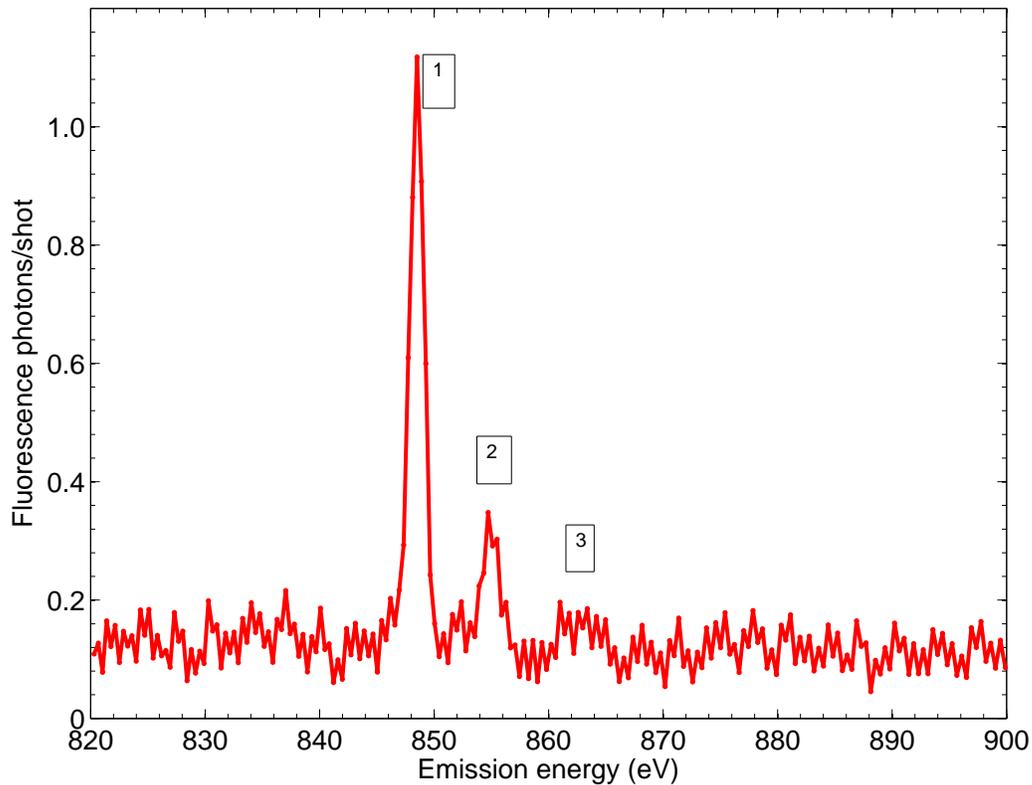}
	\caption{Neon fluorescence spectrum with three peaks, as labelled, with 1203 eV ionization energy.}
	\label{ne_plot_three_peaks}
\end{figure*} 

The raw signal from the camera is processed by summing up the pixels along the grating direction. The background electronic noise is estimated using a regression based baseline approximation method \cite{baseline_sub} and subtracted off from the original signal. The total signal is then normalized by the number of frames times FEL shots to provide the fluorescence photon counts per shot. The resulting neon fluorescence spectrum is shown in Figure \ref{ne_plot_three_peaks}.

\begin{figure*}[t]
	\centering
	\includegraphics[width = 0.5 \textwidth, scale = 0.5 ]{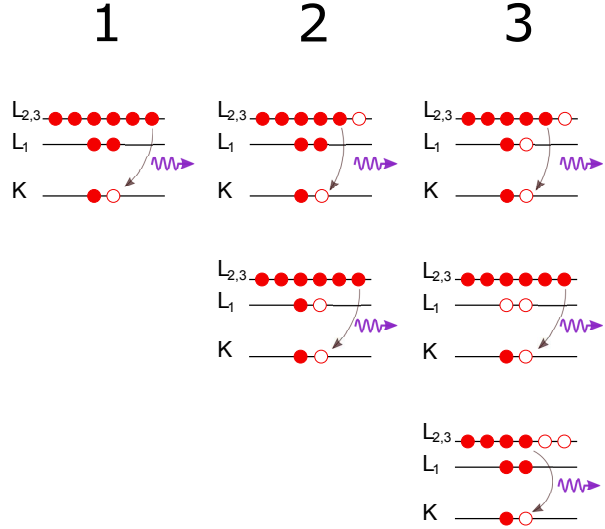}
	\caption{Electronic configuration for different pathways for neon fluorescence subsequent to core ionization. Peak 1 results from 1s$^1$2s$^2$2p$^6$ $\rightarrow$   1s$^2$2s$^2$2p$^5$ transition. Peak 2 results from 1s$^1$2s$^2$2p$^5$  $\rightarrow$  1s$^2$2s$^2$2p$^4$ and 1s$^1$2s$^1$2p$^6$ $\rightarrow$ 1s$^2$2s$^1$2p$^5$. Peak 3 is predominantly due to radiative decay after CAC from 1s$^1$2s$^1$2p$^5$ $\rightarrow$  1s$^2$2s$^1$2p$^4$, 1s$^1$2s$^0$2p$^6$ $\rightarrow$  1s$^2$2s$^0$2p$^5$, and 1s$^1$2s$^2$2p$^4$ $\rightarrow$ 1s$^2$2s$^2$2p$^3$. }
	\label{fluorescence_pathway}
\end{figure*} 

\begin{figure*}[h]
	\centering
	\includegraphics[width = 0.70 \textwidth, scale = 0.70 ]{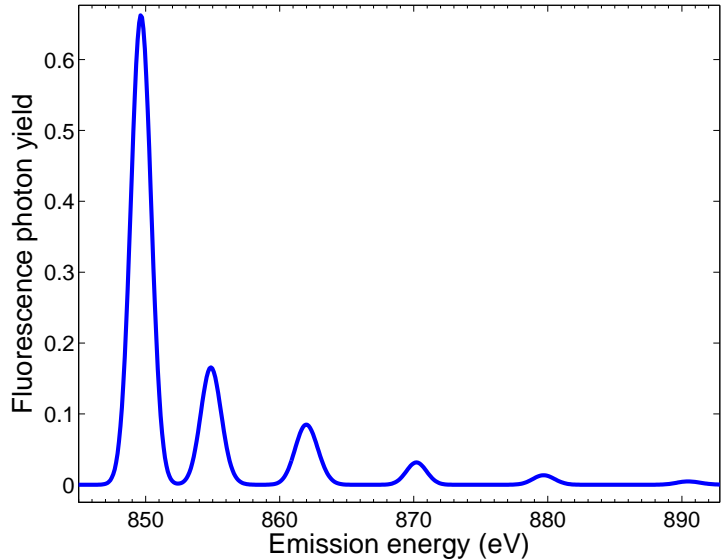}
	\caption{Theoretical photon yield of neon fluorescence using 100 fs pulse duration and 8 $\times$ 8 $\mu$m$^2$ focal spot size convoluted with a Gaussian.}
	\label{theory_yield}
\end{figure*}

Peak 1 (848.50 $\pm$ 0.75 eV) is the known peak from the Ne transition of L$_{2,3}$ $\rightarrow$ K$_\alpha$ (848.5 eV) \cite{thompson_vaughan}. This originates from the removal of a core-hole electron upon single photon absorption above resonance. The higher photon yield in Peak 1 compared to other peaks results from the Gaussian nature of the beam flux distribution causing a significant proportion of low fluence around the wings of the beam. This produces a large proportion of single core hole states which relaxes by an electronic transition from L$_{2,3}$ $\rightarrow$ K$_\alpha$. Peak 2 (854.70 $\pm$ 0.75 eV) corresponds to photoionization shake-off resulting in creation of either of 1s$^1$2s$^1$2p$^6$ or 1s$^1$2s$^2$2p$^6$ states followed by relaxation from L$_{2,3}$ to K shell, as shown in Figure \ref{fluorescence_pathway}. This photoionization shake-off lines were also observed in \cite{keski-rekhonen} using Cr-K$\alpha$1 radiation on neon gas target. Peak 3 (861.50 $\pm$ 0.75 eV) is predominantly produced from transitions resulting from the ionization produced by the two photon process of core ionization – Auger decay – core ionization (CAC) in creating hole states of  either  of 1s$^1$2s$^1$2p$^5$, 1s$^2$2s$^1$2p$^4$ and 1s$^1$2s$^2$2p$^4$ states. In \cite{keski-rekhonen}, the production of this peak is mostly due to double core ionization, with a probability of 0.0032 \cite{southword} by the Cr-K$\alpha$1 line at 5414.7 eV \cite{thompson_vaughan}. Table 1 shows the ratio of the photon number of Peak 2 and 3 relative to Peak 1, seen by their ratio of the area under the peaks, consisting of single K vacancy in the experiments in \cite{keski-rekhonen} and us. The production of 1s$^1$2s$^1$2p$^6$ or 1s$^1$2s$^2$2p$^5$ is independent of the fluence of the beam since the process is initiated by absorption of one photon followed by a shake-off photoionization \cite{saito,gao}. This is evident in the similar ratio of the area within the experimental uncertainty for $\frac{\mathrm{Peak 2}}{\mathrm{Peak 1}}$ in the two experiments. However, there are expected difference in the ratio of $\frac{\mathrm{Peak 3}}{\mathrm{Peak 1}}$ due to quadratic dependence of the CAC process on x-ray flux \cite{buth_nonlinearity}. In our experiment, we also observe the broadening of Peak 3 due to the multiplet splitting into several transition pathways producing 861.71 eV, 862.21 eV, 862.71 eV, and 870.17 eV \cite{supplementary_data}. The combination of each of these electric dipole transitions contributes in the overall shape of the peak. The  theoretical fluorescence yield is calculated by summing up the contribution of all the fine structure levels. The uncertainty in the calculation of each of the peaks’ energy is about 1 eV. The theoretical photon yield for the neon fluorescence for 8 $\times$ 8 $\mu$m$^2$ focal size is shown in Figure \ref{theory_yield}. Due to low signal-to-noise ratio in the spectrometer owing to the low target density, the other high lying calculated peaks such as 879.7 eV and 890.8 eV, producing the hole states of 1s$^1$2s$^1$2p$^3$ and 1s$^1$2s$^1$2p$^2$ respectively, are not discernible in the experiment data shown in Figure \ref{ne_plot_three_peaks}.

\begin{table}
	\caption{\label{jlab1}The ratios of area under the peak fluorescence photon yields between the experiments and theory.}
	\centering
	\footnotesize
	\begin{tabular}{@{}lcc}
		\br
		& Peak 2 / Peak 1 & Peak 3 / Peak 1\\
		\mr
		Experiment \cite{keski-rekhonen}& 0.34 $\pm$ 0.05 & 0.09 $\pm$ 0.02\\
		This experiment & 0.34 $\pm$ 0.05 & 0.22 $\pm$ 0.05\\
		Theory \cite{buth_ne} with 10.0 $\times$ 10.0 $\mu$m$^2$ focal area& 0.27 & 0.13\\
		Theory \cite{buth_ne} with 7.3 $\times$ 7.3 $\mu$m$^2$ focal area& 0.27 & 0.22\\
		Theory \cite{buth_ne} with 5.0 $\times$ 5.0 $\mu$m$^2$ focal area& 0.29 & 0.43\\
		
		\br
	\end{tabular}\\
\end{table}
\normalsize

\begin{figure*}[h]
	\centering
	\includegraphics[width = 0.70 \textwidth, scale = 0.70 ]{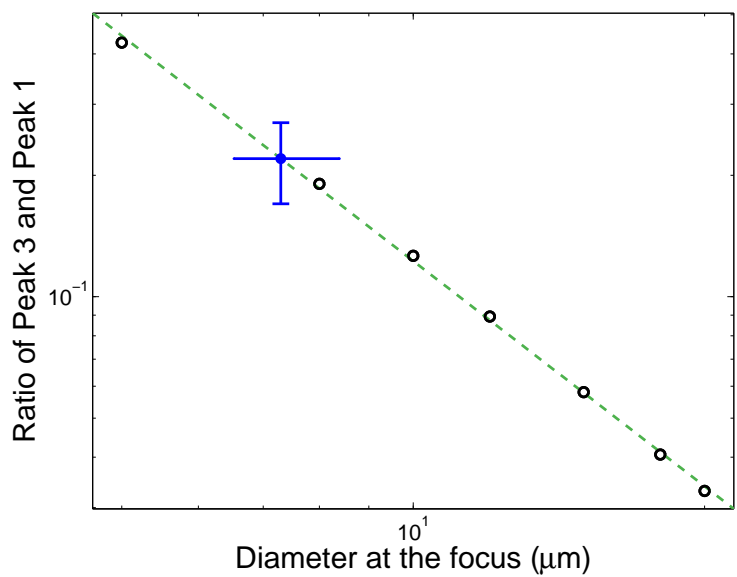}
	\caption{Ratio of area under Peak 3 and Peak 1 for different beam diameters. The black open circle is from the calculation of the fluorescence photon yield, while the green dashed line is the log-log fit. The blue filled circle with error bar is the ratio from the experiment. The slope of the log-log fit is $-1.855702$, and the intercept is 0.944305. This amounts to near-quadratic dependence of the CAC process on the beam diameter.}
	\label{ratioP3P1}
\end{figure*}

The theoretical fluorescence photon yield ratio of Peak 3 and Peak 1 for different focal size shows a linear behavior to log-log fit.  Using the photon yield ratio of those peaks found from the experiment, we can also determine the focal spot size, as shown in Figure \ref{ratioP3P1}, which in our case is $7.30_{-0.77}^{+1.09} \times 7.30_{-0.77}^{+1.09}$  $\mu$m$^2$. Most experiments infer the focal size based on the ion yield measurements of either neon or argon using calibration data that are well known \cite{youngL,motomura} or using imprints by x-rays on solid samples \cite{chalupsky}. Here, we used the fluorescence data with our theory to extract the focal spot size.

\section{Conclusion}
We demonstrated the usage of a VLS spectrometer to measure the fluorescence of neon upon 1200 eV, 100 fs FEL pulse ionization. We have demonstrated that we can use the fluorescence peak ratio to estimate the focal spot size. Upon comparing our results with previously explored fluorescence experiment of neon, we show that our results agree with the findings and provides an explanation of one of the peaks that was not identified previously in any of the experiments. We also show the viability of fluorescence measurement as a tool to provide a complementary picture to the mostly used ion electron yield measurements at FEL.

\ack
This work was funded by the Department of Energy, Office of Science, Basic Energy Sciences (BES), Division of Chemical Sciences, Geosciences, and Biosciences under grant No. DE-SC0012376 and by a DOE-BES SISGR grant No. DE-SC0002004 to UConn.  Use of the Linac Coherent Light Source (LCLS), SLAC National Accelerator Laboratory, is supported by the U.S. Department of Energy, Office of Science, Office of Basic Energy Sciences under Contract No. DE-AC02-76SF00515. SF acknowledges support by the German Federal Ministry of Education and Research (BMBF) under Contract No. 05K16SJA. We thank all of the LCLS staffs and engineers who contributed in building the soft x-ray spectrometer. 

\section*{References}
\bibliography{bib_IOP}

\end{document}